\documentclass[english, 12 pt]{article}
\usepackage[T1]{fontenc}
\usepackage[latin9]{inputenc}
\usepackage{graphicx}
\usepackage{caption}
\usepackage{subcaption}
\usepackage{babel}
\usepackage{url}
\usepackage{multicol}
\usepackage{amsmath}
\usepackage{rotating}           % for sideways tables/figures
\usepackage{color}

\makeatletter

\newcommand{\fig}[1]{Figure(\ref{#1})}
\newcommand{\tab}[1]{Table(\ref{#1})}
\newcommand{\eq}[1]{Eq.(\ref{#1})}

\newcommand{\la}[1]{ \label{#1}}
\renewcommand{\a}{\alpha}

\newcommand{\bsubs}{\begin{subequations}}
\newcommand{\esubs}{\end{subequations}}

\providecommand{\ba}{\begin{align}}
\providecommand{\ea}{\end{align}}

\usepackage{multirow}
\newcommand{\be}{\begin{equation}}
\newcommand{\ee}{\end{equation}}

\newcommand{\bea}{\begin{eqnarray}}
\newcommand{\eea}{\end{eqnarray}}

\begin{document}

\title{ Innovations in Statistical Physics V1.2\\
}

 \author{ Leo P. Kadanoff\footnote{e-mail:  lkadanoff@gmail.com}~
 \\
\\
 The James Franck Institute\\
The University of Chicago
\\ Chicago, IL USA 60637
\\ 
\\}

\maketitle

\begin{abstract}
In 1963-71, a group of people, myself included,  formulated and perfected a new approach to physics problems, which eventually came to be known under the names of scaling, universality, and renormalization.  This work formed the basis of a wide variety of theories ranging from its starting point in critical phenomena,  moving out to particle physics and relativity and then into economics and biology. 
This work was of transcendental beauty and of considerable intellectual importance.
 
This left me with a personal problem.  What next?  Constructing the answer to that question would dominate the next 45 years of my professional life.  I would
\begin{itemize}
\item Try to help in finding and constructing new fields of science.
\item Do research and give talks on science/society borderline.
\item Provide constructive criticism of scientific and technical work.
\item Help students and younger scientists.
\item Demonstrate  scientific  leadership.
\end{itemize}

\end{abstract}

\newpage

\tableofcontents
\newpage
\renewcommand{\theequation}{\arabic{equation}}
\setcounter{equation}{0}
\section{Introduction}
This is a reconstruction of a talk that I gave at the Institute of Physics, London, on January 13, 2011.   It is reconstructed to make the original   power-point style presentation suitable for book-style publication.  This  talk marked a particularly important occasion for me.  The Institute was honoring me with its Newton medal for my accomplishments in physics. The actual talk was given on the day before my 74th birthday.  Not only were there many distinguished scientists in the audience, in addition my wife and I had invited friends and relatives including my children and grandchildren (ages 7 to 20).   I had the difficult job of trying to say things that might catch the attention of all.   

\begin{figure}[!]
%\begin{centering}
%\begin{multicols}{2}
%\begin{center}
\hspace{-20
 pt}
\includegraphics[width=32
 cm ]{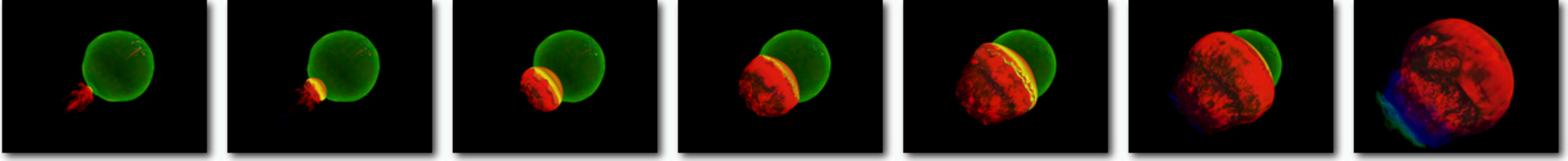}
%\end{center}
\caption{Stellar explosion.  Computer simulation of a sequence of events that will bring about a huge explosion of a star.  This kind of event is called a supernova.  I shall return to this simulation toward the end of the talk.
  }
\la{Explode}
\end{figure}

I cheated a bit by starting off with a video of a simulation of an star exploding and becoming a supernova.  (See \fig{Explode}.) Even the 7-year-old looked up.   In this context, the video was not entirely a cheat since I had been a (minor) part of the team that had constructed it and had expended considerable effort in explaining the scientific significance of the work to audiences around the world.

 I had some difficulty both in constructing the London talk and putting together this piece.   In London, my quite-varied audience wanted to know what I had really accomplished and whether I  was worthy of this high honor.  For this note, I was asked to talk about myself.  So I have here stressed my own work.   However, it is clear to me, as a student of the history of science, that every advance is the product of many minds some foreshadowing, some constructing, some extending,  some explaining. In this note,  I have played down the richness of the scientific construction process and focused on just one actor, myself.    A more extended view of the historical process is available  in Cyril Domb's book on the critical point\cite{Domb} and in my articles on the role of renormalization in the history of science\cite{LPKHist1,LPKHist2,LPKHist3}. 

\section{My accomplishment (1966-68)
}

I started out in science in times much easier than our own.  Sputnik had produced a flurry of support for science so that there was  a temporary shortage of scientists.  I went through Harvard  very well supported by the NSF and jobs in the missile industry, and moved to an NSF postdoc in Copenhagen in which I was better paid than the local professors.   I never had to look for a job.  Through the kind offices of David Pines and John Bardeen a good faculty job in Urbana came to me.  Despite, or perhaps because, of all this I had by the time my story starts in 1966, a set of quite respectable (but conventional) physics accomplishments.

Like every young scientist, I dreamt of doing something really new, ideally in a short paper, something that would have some impact upon science.  A sabbatical in England had fallen into my lap and I used that time to get into a new (for me) scientific area:   phase transitions.

     In the early 1960s, the behavior of materials near the critical points of phase transitions was a quite fashionable subject.    Science pretty well understood so-called {\em first order phase transitions}, as for example the boiling of water. This kind of transition involves a sudden change in material properties,   like the jump in density upon  boiling water.  But a temperature adjustment may reduce the size of the density-jump.  When the jump becomes infinitesimally  small,  and the phases become almost indistinguishable the system is said to be approaching its {\em critical point}. In this region of the phase diagram, the material's behavior becomes harder to understand.   While approaching criticality, thermodynamic derivatives tend toward infinity and one can see large patches of the material  deviating from average behavior.   These novel effects suggest questions about the fundamentals of statistical mechanics and the dynamics of materials.  These effects called for a theory.

Consequently, along with others,\footnote{ The most important work came from Lars Onsager\cite{Onsager}, Cyril Domb\cite{Domb}, Michael Fisher\cite{Wilson-Fisher}, Benjamin Widom\cite{Widom}, A. Patashinski, V. Pokrovski\cite{PP},  and (most crucially) Kenneth Wilson\cite{RG1,RG2,RG3}. } I was trying to understand the behavior of condensed matter systems near critical points.    In England, I had done a lengthy and detailed calculation of fluctuations near the transition of  the two-dimensional Ising model. This model provided a simplified description of a phase transition and a critical point.   I was trying to understand the meaning of my work.  It would turn out that my calculation would provide an important clue to the behavior of all kinds of materials near their critical points. But, since the Cavendish Lab could not afford to send  preprints from England, and because the refereeing process involved many iterations of slow mail over oceans, I was almost the only one who knew the contents of my lengthy calculation.

\begin{figure}[!]
 \includegraphics[width=0.4\textwidth]{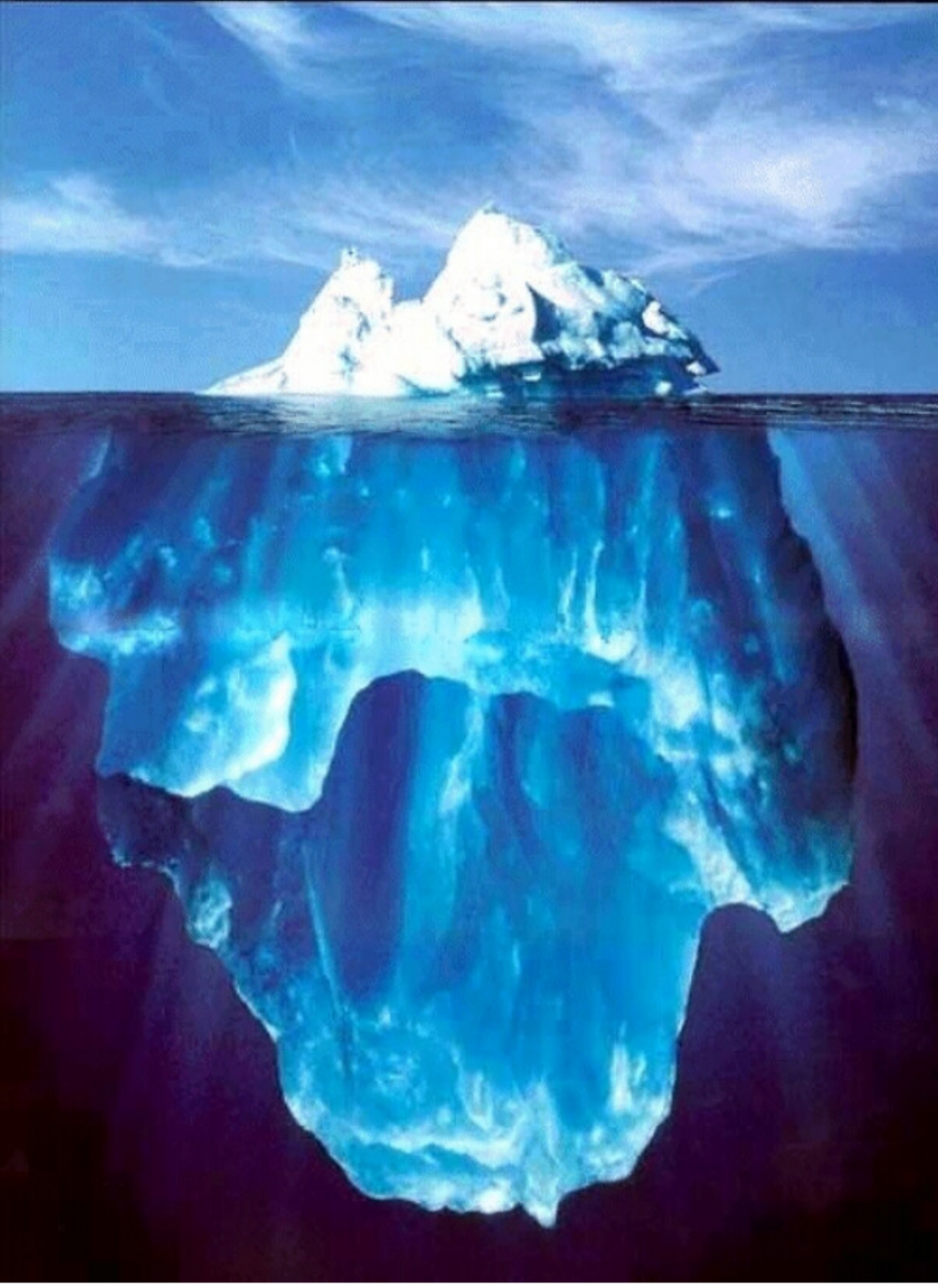}

\caption{ The phases of water exhibited in an iceberg (solid), floating in the ocean (liquid), surrounded by air containing water vapor (gas).   The familiar world is filled with materials and their different phases.
  }
\la{iceberg}
\end{figure}

\begin{figure}
         \centering
         \begin{subfigure}[b]{0.35\textwidth}
                 \includegraphics[width=\textwidth]{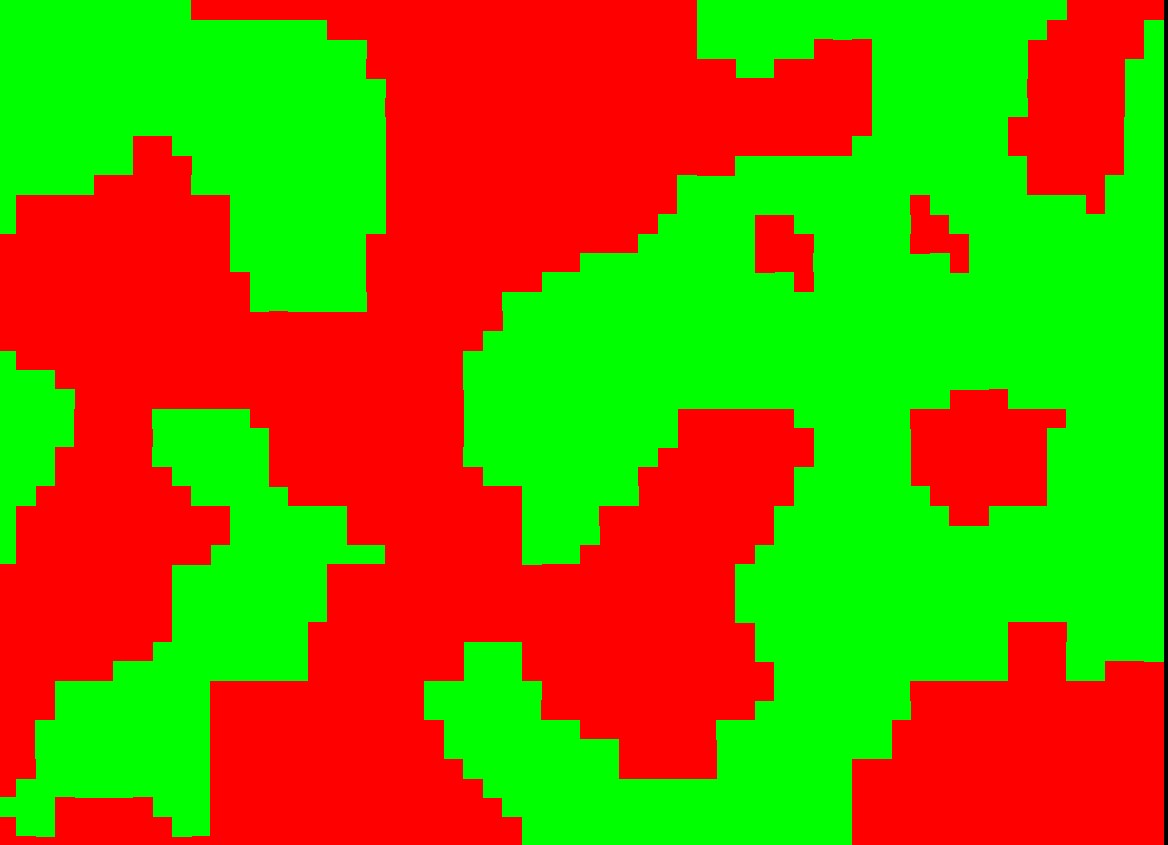}
                 \caption{Two phases touching}
                 \label{Boiling}
         \end{subfigure}%
         ~ %add desired spacing between images, e. g. ~, \quad, \qquad etc.
         \begin{subfigure}[b]{0.35\textwidth}
                 \includegraphics[width=\textwidth]{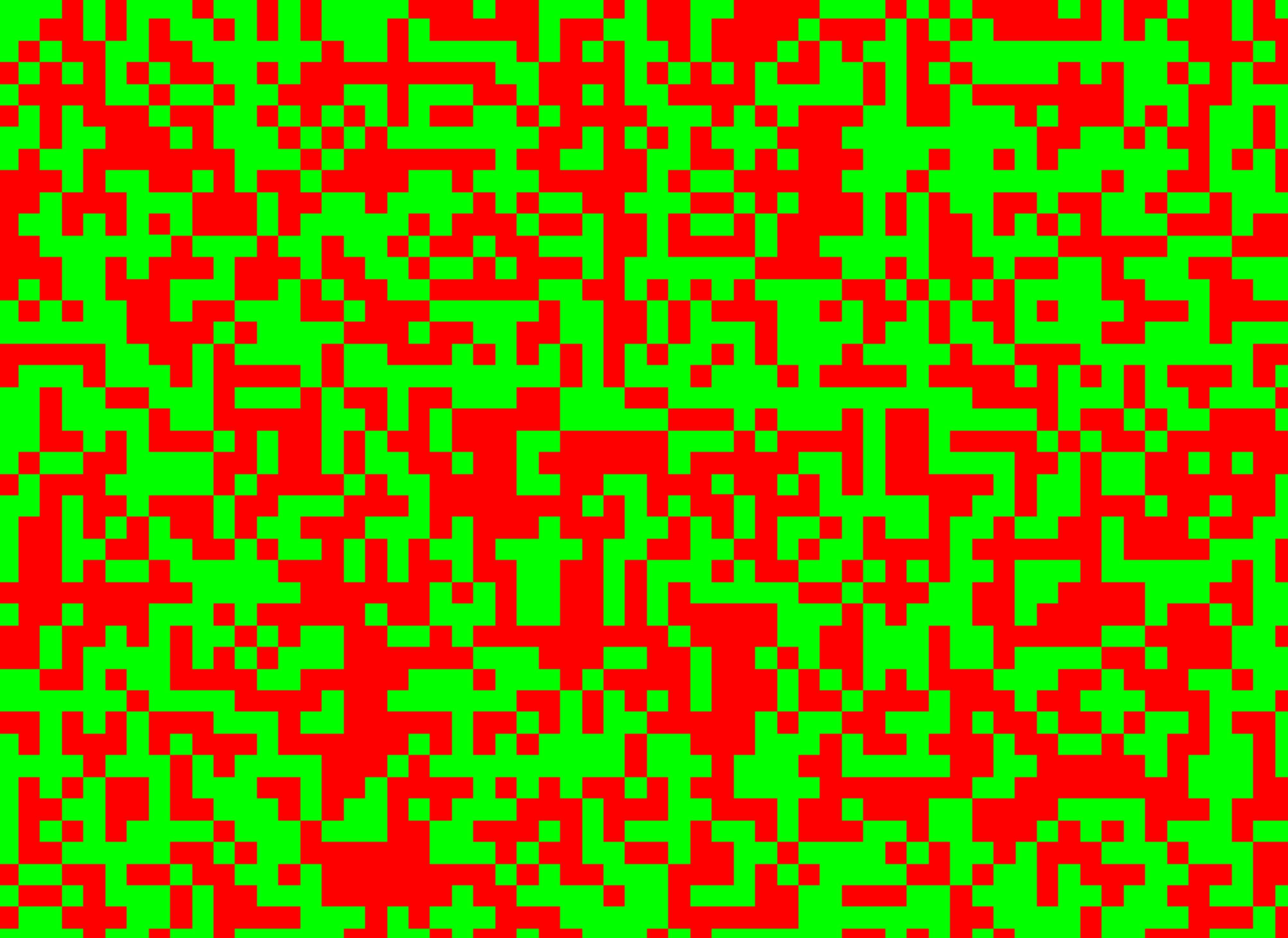}
                 \caption{High Temperature}
                 \label{HighT}
         \end{subfigure}
         ~ %add desired spacing between images, e. g. ~, \quad, \qquad etc.
         \begin{subfigure}[b]{0.36\textwidth}
                 \includegraphics[width=\textwidth]{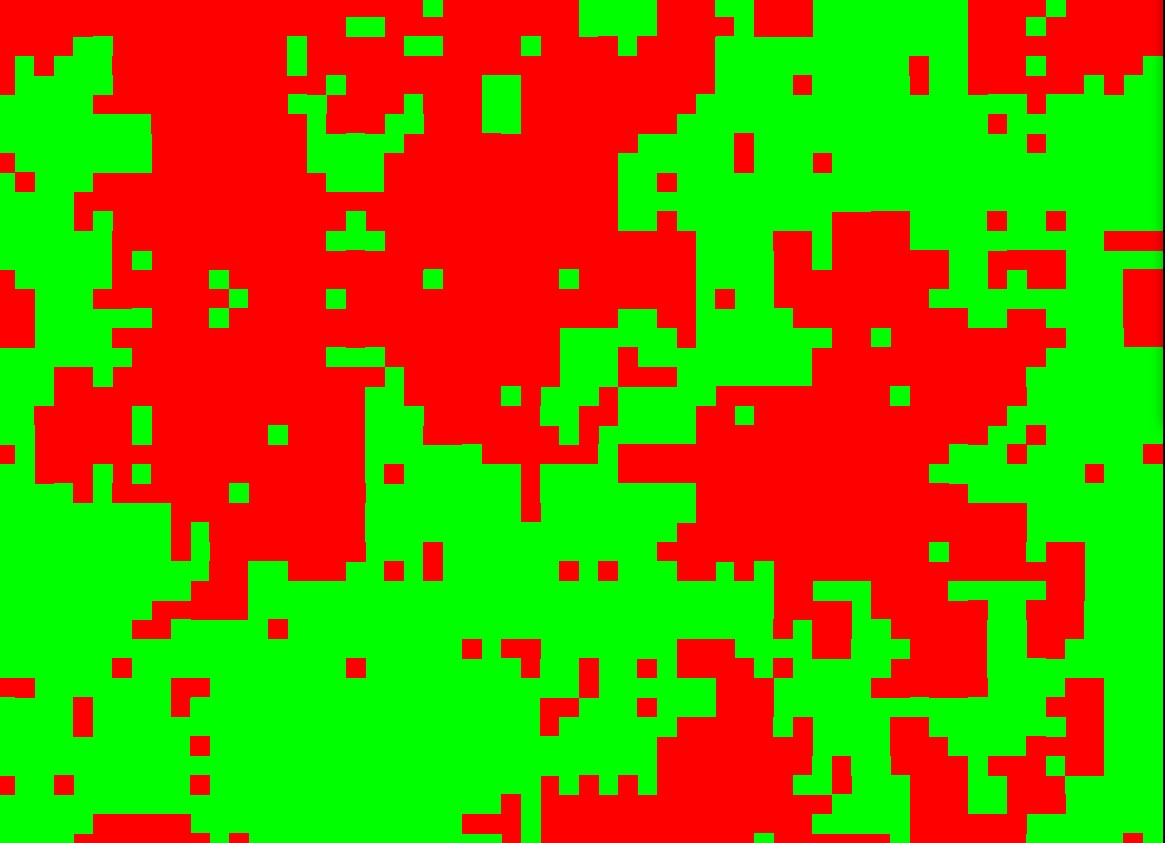}
                 \caption{Critical point}
                 \label{Critical}
         \end{subfigure}
         \caption{Phases of water as exhibited in the Ising model.   At low temperatures, in  plate (a), we can see large high density and low density regions in contact with one another.    At high temperatures, as in the next plate, all densities are intermixed in a fine-grained way.  At the critical point,  plate (c), unmixed regions of all sizes occur.   }\label{IsingPhases}
\end{figure}

The Ising model\footnote{It is named after Ernst Ising,  who first studied it.}  is a simplified view of a fluid\cite{Yang-Lee}, and provides a rough, but qualitatively accurate,  description of an actual fluid. The Ising fluid is divided into boxes of two kinds, high density and low density. The  forces within the fluid tend to pull high density regions toward other high density regions and thereby leave islands of low density regions. (See \fig{IsingPhases}.)  At lower temperatures these forces can hold the different regions together, forming huge islands composed of a single phase,  so at boiling the fluid looks like what you see in \fig{Boiling}.  At high temperatures, the fluid is all shaken up and the different regions do not hold together so that red and green intermix as in \fig{HighT}.   My own interest turns mostly to the situation in between high temperatures and low.  Here, islands form of red regions and green regions.  At just the right temperature, called the {\em critical temperature},  red and green islands of all sizes form and intermix as in \fig{Critical}.   

In what follows, it will be important to recognize the two parameters that control this system's behavior near criticality.  I just mentioned the first, the temperature.  The second parameter is the pressure, which is the primary determinant of the proportion of high density regions relative to the low density ones.  This is included in the analysis as $p$, the pressure minus the critical pressure.  Both parameters are zero at the critical point. 

An important empirical fact was supported by my long calculation: At each temperature, there is a characteristic size of the fluctuating regions, a size which grows as criticality is approached. The characteristic size, $\xi$, grows as a power of $t$, a dimensionless measure of the temperature deviation from criticality.  The equation is
\be
\xi=a t^{-\nu}
\ee    
where $a$ is a length which  roughly gives the range of the microscopic forces that drive the transition.   The number, $\nu$, is an example of an important quantity, called a {\em critical index,}  that we would like to evaluate. The transition is characterized by a large number of diverging statistical or thermodynamic quantities, each with its own characteristic critical index.  

Back to my story:  Because of my calculation, and my knowledge of what others have done,  I had a reasonably extensive acquaintance with the behavior of the two dimensional Ising model.   But, all this knowledge is in disconnected pieces and I could not make it into a meaningful picture. 

Then, in one week at Christmas time in 1965, I see how to bring all these pieces together\dots 

\subsection{My insight}  
As I said,  at the critical point  the Ising system contains  regions of all sizes.  How can we analyze this situation?   Maybe we should use different descriptions based upon viewing the Ising system at  different magnifications.  We know that the basic atomic constituents of water are very small, so there is plenty of room for magnification between their size and our own.  But how to change the magnification of our view. Well that is the new insight which, forty-five years later,  brought me to London.  

\begin{figure}
         \centering
         \begin{subfigure}[b]{0.3\textwidth}
                 \includegraphics[width=\textwidth]{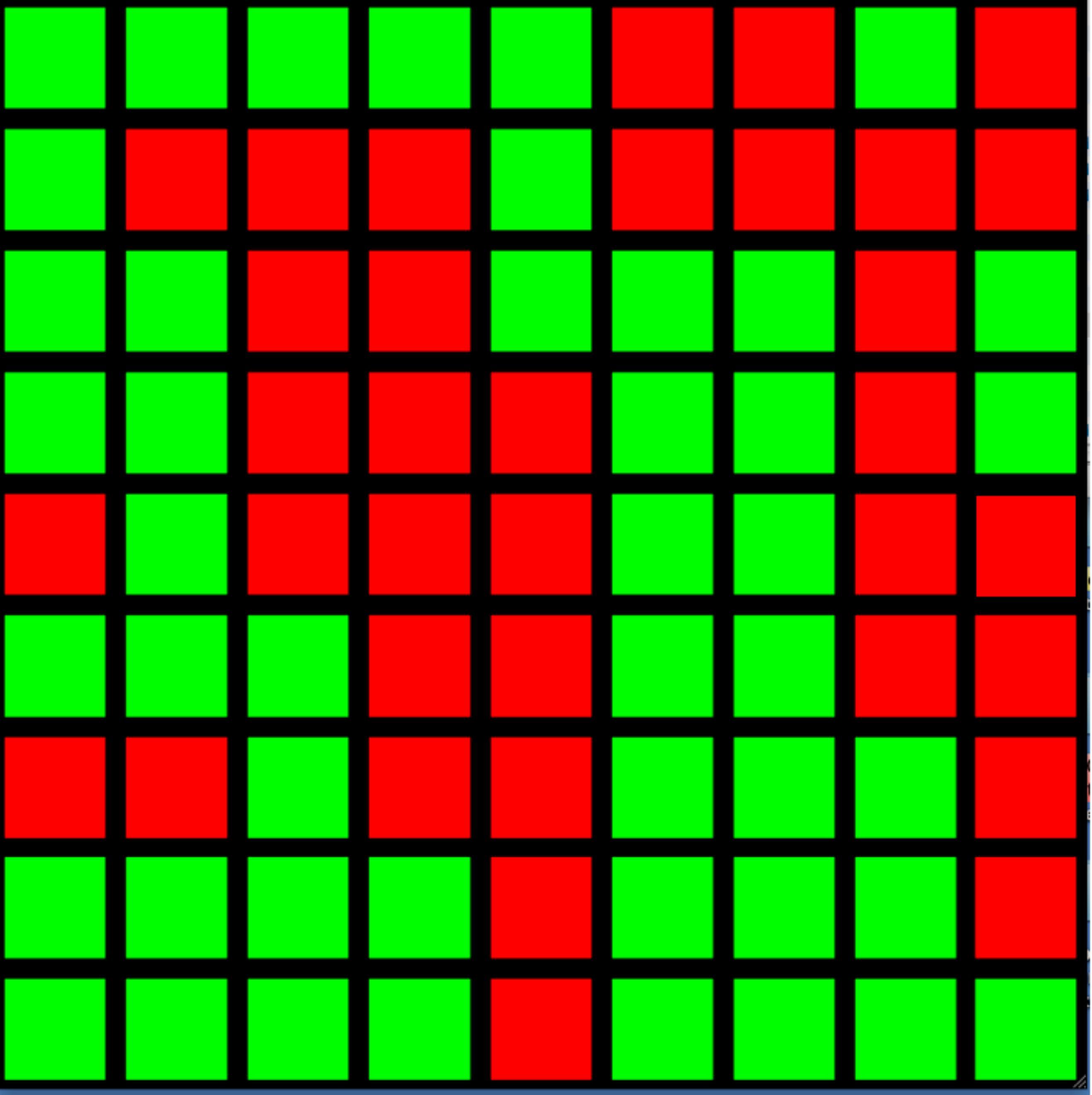}
                 \caption{Ising criticality}
                 \label{Iboxes}
         \end{subfigure}%
         ~ %add desired spacing between images, e. g. ~, \quad, \qquad etc.
         \begin{subfigure}[b]{0.3\textwidth}
                 \includegraphics[width=\textwidth]{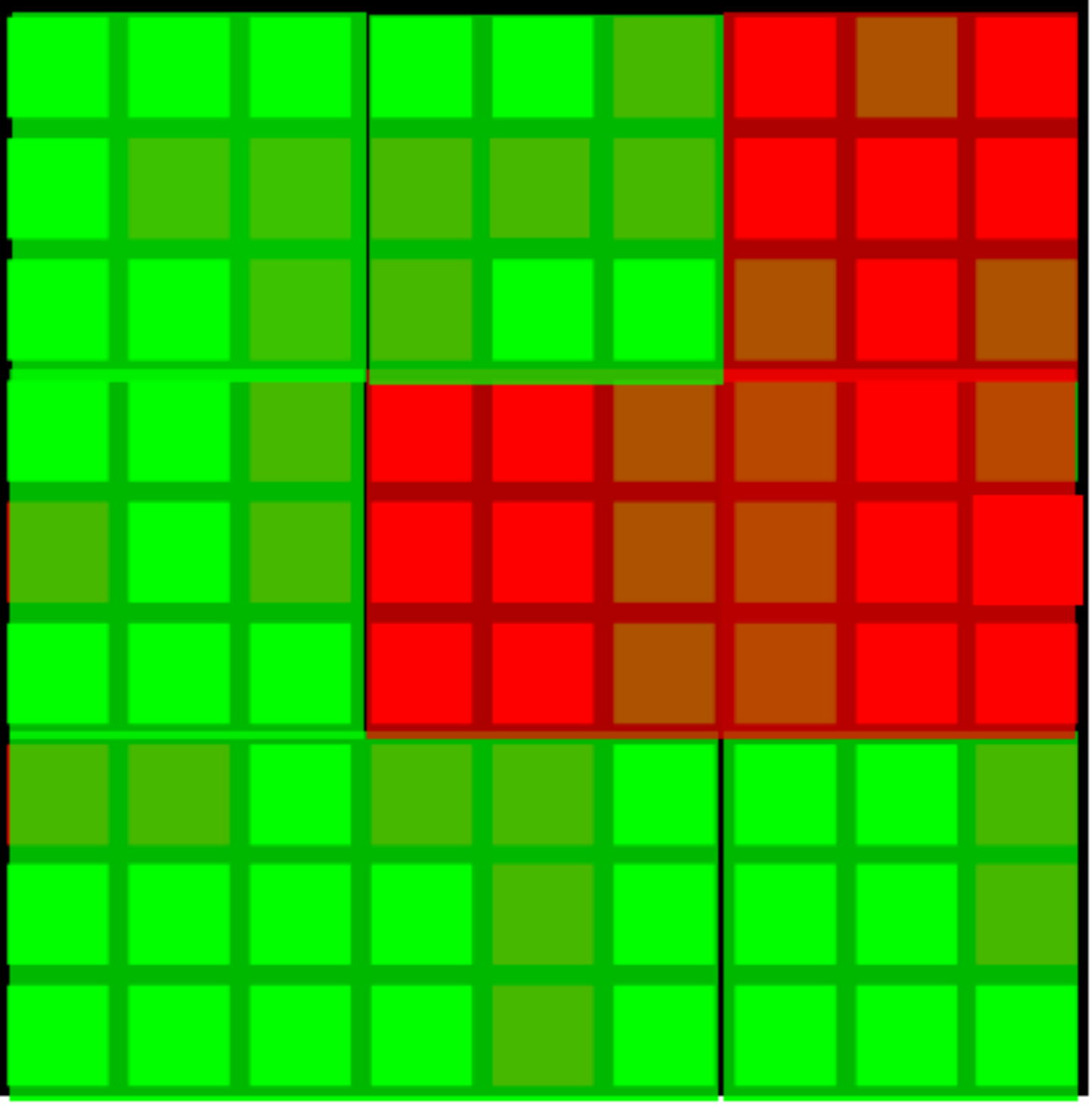}
                 \caption{Box construction}
                 \label{coveredBoxes}
         \end{subfigure}
%         ~ %add desired spacing between images, e. g. ~, \quad, \qquad etc.
%           %(or a blank line to force the subfigure onto a new line)
         \begin{subfigure}[b]{0.3\textwidth}
                 \includegraphics[width=\textwidth]{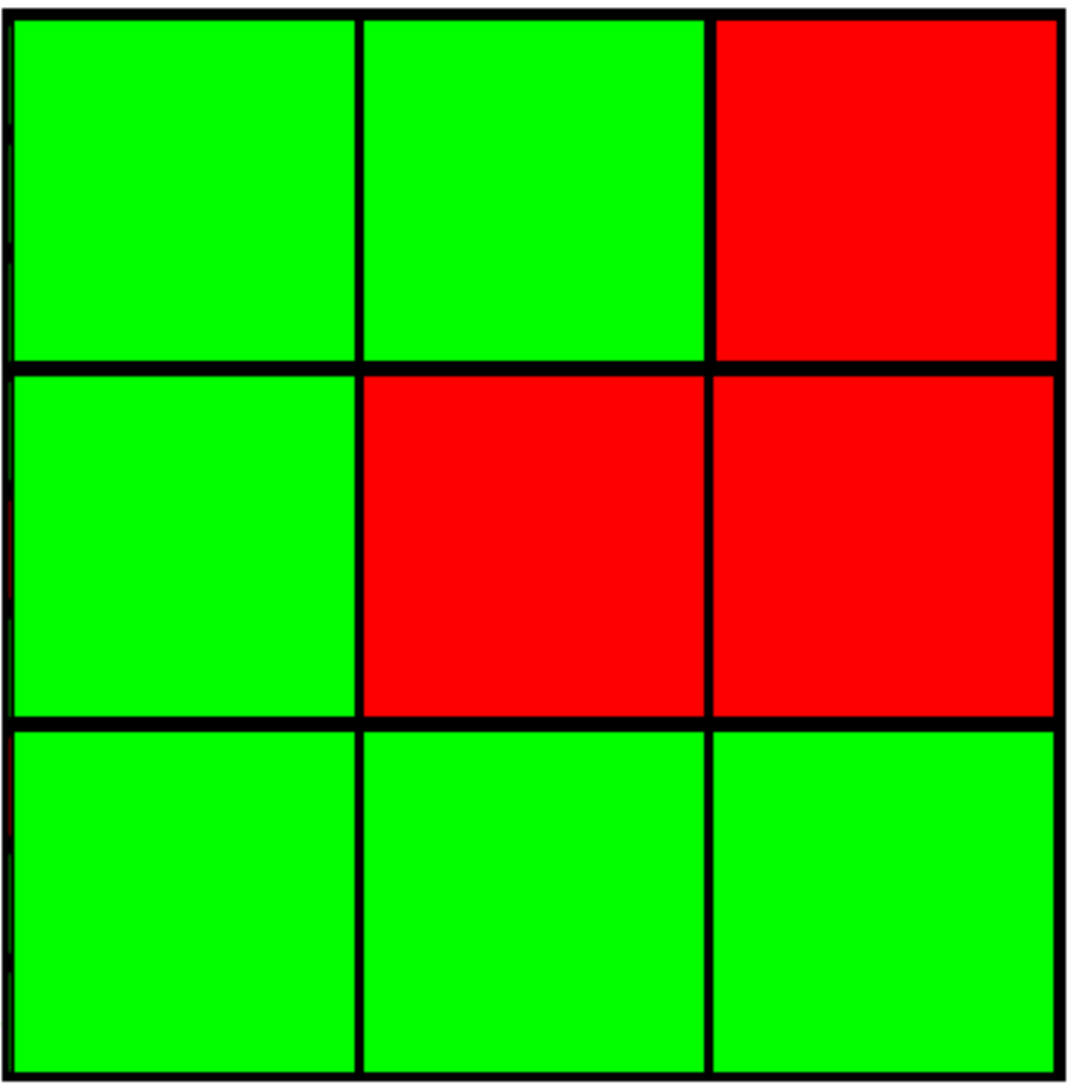}
                 \caption{A new magnification}
                 \label{vapor}
         \end{subfigure}
         \caption{A renormalization calculation in the Ising model. The first panel shows a configuration of the model with high density boxes in red.  In the second panel, we have covered the system with bigger boxes, each with a side  three times larger than in the first panel.  These bigger boxes are colored according to the ``majority rule'' in which a bigger box is colored  to agree with the majority of boxes beneath it.  The final panel shows the renormalized model, with bigger colored boxes.  }\label{renormalization}
\end{figure}

Look at \fig{renormalization}, please.  The first panel shows a small region of the Ising model at its critical point with its mix of high density and low density squares.  In the second panel, \fig{coveredBoxes}, we have divided  this fragment of fluid into three by three boxes.  Nine boxes cover the whole panel.   These bigger boxes each have a color.  If the majority of the small  covered squares are red we give the covering box a red color.  Otherwise, it gets a green color.   The right hand panel shows just the covering boxes. That panel looks like a more coarse version of the critical Ising model.      

I then made the hypothesis, now called a {\em universality hypothesis,} that at criticality very much the same description will tell you about both the new box problem and the original Ising model. This hypothesis is based in the experimental observation that different critical phenomena problems have similar observable behavior.  I went further and assumed that they were really the same so that exactly the same description could work for problems on different length scales.   In this form, my assumption was one of  {\em scaling}, i.e. that all feature of criticality would remain unchanged as one changed the length scale from that of the Ising model to that of the box.

Instead of solving either the original Ising model or the box problem, I talked about how to relate the two problems.  The usual way of describing liquid-gas criticality is in terms of the two already-mentioned measures of how far the actual system is from the critical point, $t$, a dimensionless measure of the temperature deviation from criticality, and $p$, a similarly dimensionless measure of the  pressure deviation\footnote{For example we could define $t=1-T_c/T$, where $T$ is the absolute temperature and $T_c$ is the critical temperature.}.

In this way, I could describe  the same physical system in two different ways.  The old description was of $N$ squares, with size $a$, and thermodynamic variables $t$ and $p$.    The new, box-based, description  is just the same except that the parameters have different values.  These might be written as $t_{new}$ and $p_{new}$. These values have changed because we have increased the size, $a$,  of the basic box by a factor of $l$.  (In the figure, $l=3$.) Thus, the box size is
$$a_{new} =l a $$
while the number of boxes is
$$N_{new} =N/l^2 $$ since the system is  in two dimensions. 
The important new hypothesis is that in this situation the two measures of the deviation from criticality vary according to 
$$t_{new} = q~t   $$
$$p_{new} =r~p $$
where $q$ and $r$ depend upon the length change, $l$.   Because physical quantities, like $\xi$, tend to vary as powers of $t$ and $p$, I could guess that I should express $q$ and $r$ as powers of $l$.  As we shall see in a moment, this choice  enabled the elimination of $l $ from physical results. Thus, I defined
$$  q=l^{y}   $$ 
$$  r=l^{z}  $$
In 1965, $y$ and $z$ function were unknown numbers.

This transformation provided a description of the effect of scale changes upon the Ising model in terms of the numbers $y$ and $z$.  All the qualitative properties  of the phase transition can be expressed in terms of these two numbers.  For example, from what we have so far, the typical fluctuation size, called the coherence length has the value in the ``new'' picture.
\be   \la{scaleXi}
\xi_{new}= a_{new} (t_{new})^{-\nu}  =a~l~ (t l^y) ^{-\nu}      =a t^{-\nu} l^{1-\nu y}.
\ee
 But $\xi$ is a physical quantity which does not change with our representation of the system.   Therefore, all the factors of $l$ must cancel out of the right hand side of \eq{scaleXi} so that
$$   \nu=1/y. $$
The physically significant index, $\nu $ is thus expressible in terms of our model quantity, $y$.  In fact all the qualitative properties of the transition can thus be expressed in terms of $y$ and $z$.   

If we could but find $y$ and $z$\dots

\subsection{Next steps}
I gave seminars about this work at Urbana, Cambridge, Paris, Moscow, and many other places. Almost everyone loved the work, and encouraged me.  However, somewhat incongruously, for six years nobody could put the finishing touches upon it. So what could I do?  Like everyone else I was stymied.  However the Ising model  was the site of my big achievement.  Again and again I tried to find $y$ and $z$, but instead went in circles, making no progress at all.

Then, in 1971, Kenneth G. Wilson showed how to marry this box analysis with the earlier work of Murray Gell-Mann and Francis Low\cite{G-ML} and thereby produce a complete theory.  Wilson called the outcome,  using the name from the earlier work, {\em the renormalization group theory.}   

Given the fullness of time, and 1971 was forty-three years ago, we can appreciate the power and impact of  this work and of the ideas that came out of it.   

\subsection{New ideas}
\begin{itemize}
\item   {\bf Scaling.} In physics we are often concerned with connecting problems at different length scales. For example, we start from atomic forces, based upon phenomena at $10^{-10}$ meters and use these to predict the properties of solids, including scales up to many meters.     The renormalization method enables one to connect the physics at different scales, thereby understanding the relationship among different domains of physics.

\item {\bf Universality.}  As one traverses the different length scales, one only retains a few characteristics of the original problem.  Thus many different microscopic problems have the same macroscopic manifestation, depending mostly upon the symmetry of the original problem.  For example, the Ising model, a magnetic solid,  and a liquid-gas phase transition all have the same behavior near their respective critical poinst.  

\item {\bf Universality Classes.}  Problems then fall into a few different categories depending upon the nature of their solutions.  These categories are called {\em universality classes.}  It is now commonplace to say that critical behaviors fall in a few universality classes.  Moreover, scientists now use this phrase to describe many other kinds of things beyond phase transitions.  One of these days I expect to hear about a description of pop music in terms of universality classes.
 
\item {\bf  Renormalization.}  Wilson described the effect of many  successive renormalizations in which the length scale was changed again and again, as the resulting motion of the interactions toward a {\em fixed point,}  a state of constant and unchanging interaction.  In doing this, he  provided a calculational framework from which one can derive scaling, universality, and the values of the critical indices. In this way, he tied all aspects of the theory into a coherent whole.

\item {\bf New Calculational Paradigm.} Previously one started with a ``problem,'' i.e. a  description of the interactions among the particles in the system.  The job of the theorist  was mostly to calculate an ``answer'',  that is a detailed description of the the behavior of the particles.  

In the new era, one does renormalization calculations in which one starts with one problem (i.e. a set of interactions) and constructs another, equivalent, problem by changing the length scale.   A partial, but often sufficient, description of the behavior is encoded in the relation between the different scales.  
\end{itemize}

\subsubsection{Example:}
Old era:  One said that quantum electrodynamics was characterized by an interaction strength called $\a$, having the value 1/137.036.

New era:  One says that at smaller distance scales, $\a$ gets larger and eventually this electromagnetic interaction becomes very strong.  In informal language, one says that the interaction strength runs.
 
The practice of physics has changed and moved away from the mode of calculation of Newton, Boltzmann, Einstein, and Dirac, going from solving problems to discussing the relationship among problems.  Of course, this change is not all my doing.  In addition to the aforementioned people (See footnote number 1.) who pushed in this direction, all the workers in the field determine the subject's contents.

\section{Afterwards\dots}
So here I am in 1971.  Wilson had just put the beautiful finishing touches upon the theory that I partially developed.   My initial reaction:

I am disappointed and angry.  How could someone else finish up the description of my beautiful world!   

But that view, if held too hard and strong, would have paralyzed me and prevented any future work.    I concentrated on the fact that Wilson had quite impressive skills and knowledge. 

He could apply his thought processes in ways inspired by computer programming.   

He knew about previous work in particle physics and dynamics of which I was ignorant.

He was, I could acknowledge, a very great physicist.   

I was gradually able to put aside my childish chagrin, and take pleasure in the fact that, with Wilson's additions, an edifice of transcendentally beauty and  importance had been constructed. 

\subsection{But, what next?}
It was evident to me that I had been wonderfully creative, but also wonderfully lucky.   It should have been, and perhaps was, evident that I would never do anything that approached my previous level of creation.  

History suggests that theoretical physicists hit their top level of creation early in their professional lives and then never again approach the same pinnacle.  (See \tab{AgeTable}.)

\begin{table}[t]

\begin{tabular}{ |  c  |  c | c | c  | }
\hline
\bf person & \bf age  & \bf accomplishment &\bf  year  \\
\hline \hline
 Isaac Newton &  25 & mechanics & 1667 \\
Albert Einstein & 26 &  best year  & 1905  \\
P.A.M. Dirac & 26 & Dirac equation &  1928  \\
Hans Bethe & 30  & energy from sun & 1936\\
Brian Josephson & 25 &Josephson effect & 1965\\
Leo Kadanoff &29  & scaling, universality& 1966\\
K.G. Wilson & 35 & renormalization & 1971\\
Alan Guth & 32& inflationary universe & 1979 \\
\hline \hline % \\
\bf exceptions &~ & & \\ 
\hline
Galileo Galilei & 58 & relativity &1632 \\
Neville Mott & 44 & Mott transition & 1949 \\
Edward Witten & {\em passim} & many  &\\
\hline

\end{tabular}

 \caption{Table showing the ages at which theoretical physicists constructed their most creative accomplishment, along with the accomplishment and the year in which it was achieved.  Aside from myself, the people in this table were great physicists who far outdistanced me in the quality of their work. But their experience was nonetheless probably relevant to my work life. }  \label{AgeTable}
 \end{table}

\subsection{Perspective}
The historian of science, Thomas Kuhn\cite{Kuhn} says that
most scientific work is  ``normal science,'' that is constructing a small examples or extensions of a subject in which the basic approach is already accepted. 

Occasionally the basic approach comes under attack and something really new is done.  This is ``revolutionary science.'' 
\vskip 3 in
In the 1970s, I had read Kuhn and could recognize the revolutionary aspect of my past work, but also that I was likely to have only normal science in my future.  So be it. But within this limitation I could do worthwhile things, including 
\begin{itemize}
\item  Try to exemplify a standard of good work..
\item Lead my colleagues into new areas of research.
\item Offer worthwhile commentary about the relation between science and society.
\item Offer careful criticism of  scientific work. 
\end{itemize}

\subsection{New scientific directions:  toward macroscopic physics}

For decades, physics and chemistry departments had been studying atomic and subatomic phenomena, and mostly ignoring the observable world of wind, water, waves, and sand.   Partially because the existence of computers made possible new modes of investigation, in the 1970s major opportunities arose in the scientific study of study of things that happen in the observable world around us.   Commonplace things like the flow of water or the piling up of sand. 

Albert Libchaber, Sidney Nagel and I introduced\footnote{Not quite introduced. The earlier work of Chandrashekhar and Eugene Parker prepared the University to accept work in classical physics.} the study of these things at the University of Chicago, and through Chicago into physics departments all around.   For example, I did a theoretical study of the pinching off of the neck that connected a drop to a larger body of liquid.  The mathematical content of this work included the study of singularities, infinities in fluid flow.    This led to experimental work, most notably by Sidney Nagel\footnote{For a recent example of Nagel and coworkers see\cite{Nagel}.} and additional theoretical  work at Chicago on the shape of drops of fluid.  For example, the first panel in \fig{DripFigures} shows a drop of fluid, connected to a larger mass by a thin neck, just about to separate and fall off. 

At the same time, P. de Gennes was broadening perspectives of scientists in France, as was Sam Edwards in England.   Like these other people, Chicago's workers on these topics had to overcome some conservative bias, but we persevered, and physicists at other institutions gradually began to work on macroscopic behavior.

\subsubsection{Granular materials, sand}

Along the same lines, I began a series of calculations to understand what would happen in real collisions among small pieces of material like gains of sand or of sugar\cite{LPKGran}.  This too led to experimental work at Chicago by Heinrich Jaeger and Nagel.  I show only one example of very many. The second panel in \fig{DripFigures} shows a combination of work on drops and granular materials. In the this panel, a granular material is mixed into a fluid. Then the mixture is allowed to form a drop and fall off. Some steps in the process are shown\cite{Jaeger}.   

\subsubsection{Convective turbulence}

Another thread of work is in ``convective turbulence.'' 

That sounds formidable but it is exemplified by a most common situation:  heating water to boil an egg.  You heat the water from below and cool it from above.  There is hotter water toward the bottom of the pot and cooler water toward the top.  The heated water tends to expand so the water at the bottom is less dense.  The resulting disparity in gravitational force pushes the bottom fluid upward and the top fluid downward.  A flow begins.

\begin{figure}
         \centering
         \begin{subfigure}[b]{0.3\textwidth}
                 \includegraphics[width=\textwidth]{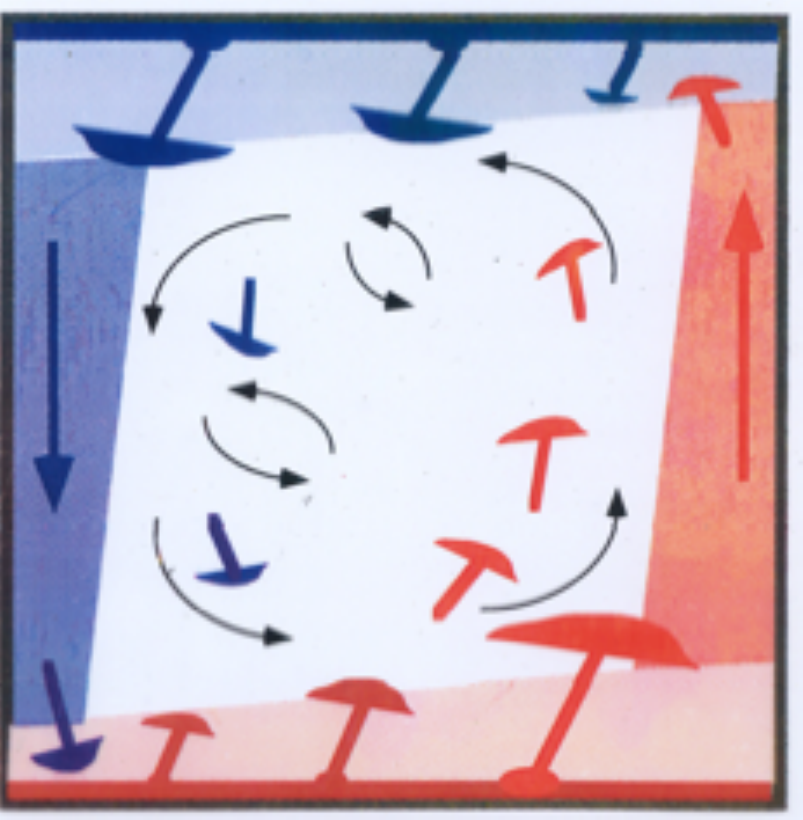}
                 \caption{Cartoon Cell}
                 \label{CartoonCell}
         \end{subfigure}%
         \begin{subfigure}[b]{0.32\textwidth}
                 \includegraphics[width=\textwidth]{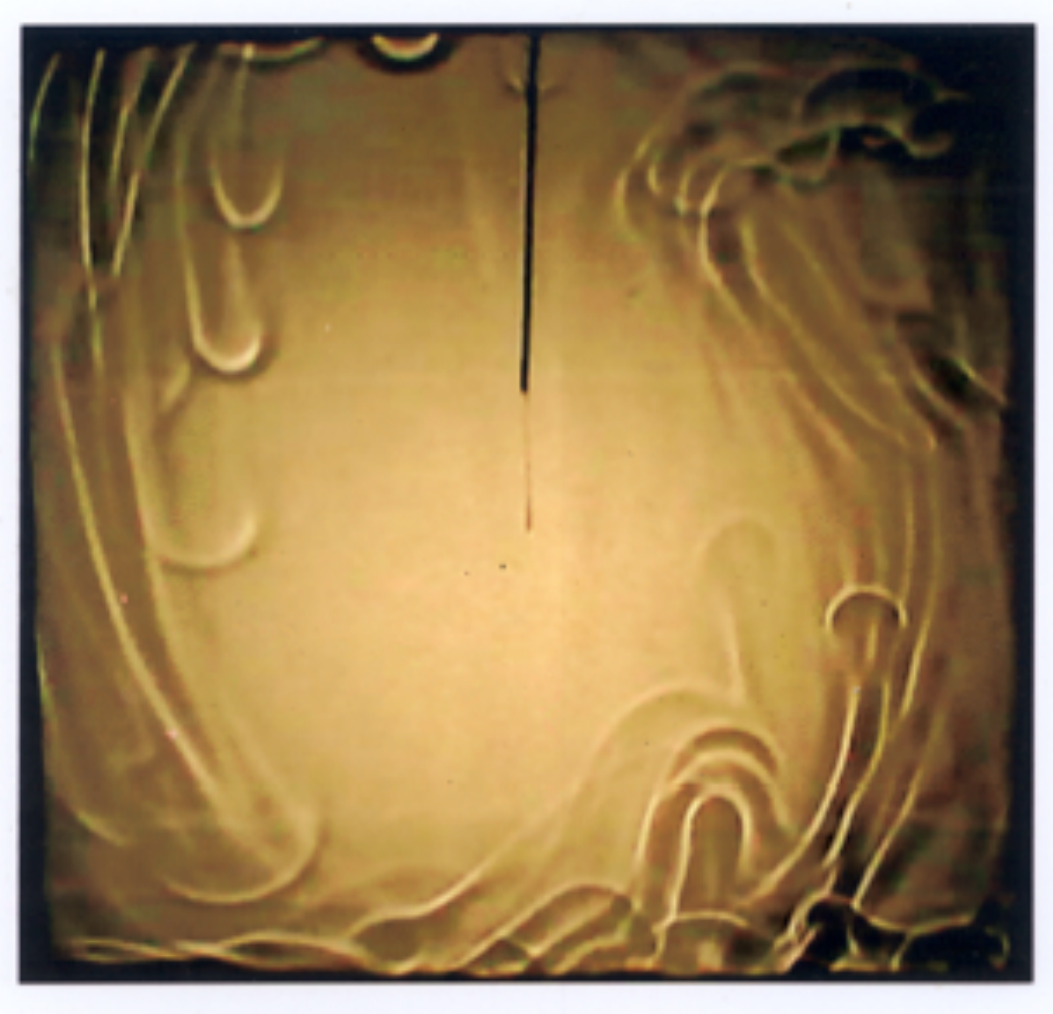}
                 \caption{Laboratory Study}
                 \label{Real Cell}
         \end{subfigure}
         \begin{subfigure}[b]{0.3\textwidth}
                 \includegraphics[width=\textwidth]{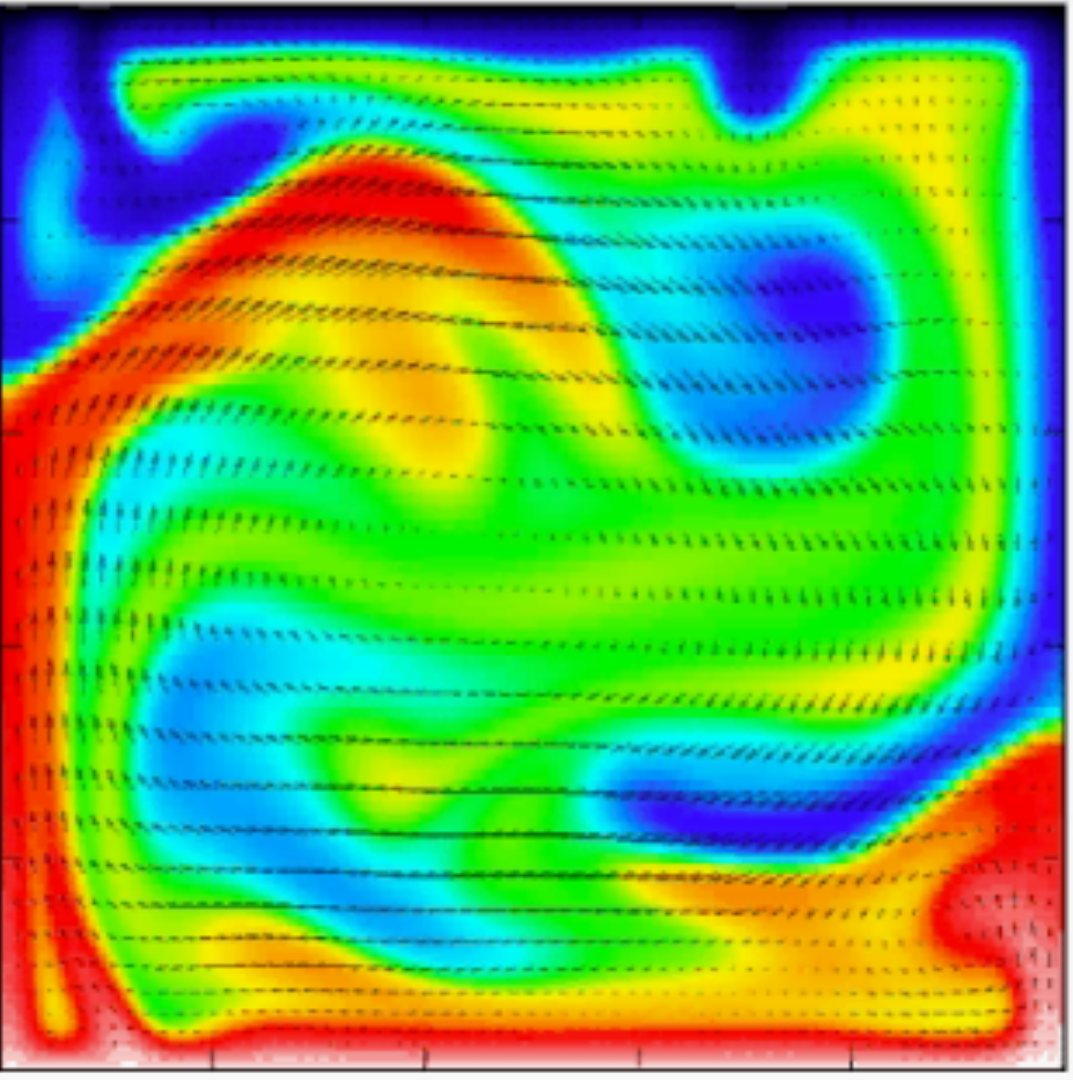}
                 \caption{Computer Study}
                 \label{Simulated Cell}
         \end{subfigure}
         \caption{Convective turbulence.  Three pictures of a box containing a fluid heated from below and cooled from above showing temperature patterns. The left-hand panel is my cartoon view, partly improved from the picture I drew before the system was actually observed in the lab. Here, and on the right, red indicates warm fluid, blue cold fluid.  The view in the center is from Penger Tong's  lab in Hong Kong. Here, shadows outline regions where the temperature changes.  The right hand panel is the result of a computer simulation by Richard Stevens, R. Verzicco, and D. Lohse. In all panels,  hotter regions are buoyant and rise, cold regions sink.  .  As you can see from all three views, this simple machine, driven by only a heater, has developed a complex pattern with mushroom-like objects called ``plumes'' in the interior and wall regions at top, bottom, left,  and right with a very different flow pattern from the center.      }\label{ConvectiveCell}
\end{figure}

The resulting motion was being investigated in the laboratory of my colleague,  Albert Libchaber. Before I saw the fluid in motion, I drew a picture of what I thought was happening.  To see my picture and the real thing, look at                            \fig{ConvectiveCell}. This shows three different views of the motion.   A cartoon picture of the convective cell shown in the first panel of the figure.   This cartoon showed different regions of the cell: jets at the side walls, thin boundary layers top and bottom, thicker mixing zones above the bottom boundary layer and below the bottom one, and a large central region containing ``plumes.''  These are mushroom-shaped objects formed in the mixing zones that can group together in the jets and move freely through the central region.  My work then described in a qualitative fashion how all these parts worked together to make an efficient ``machine-like'' fluid flow that moved heat from bottom to top in a quite efficient fashion.    I also made quantitative predictions, now outmoded;  but the qualitative picture of the flow's structure remains valid.

\subsection{The ``Intelligent Design'' discussion}
The convective turbulence result formed a major part of a series of lectures that I gave dozens of times to a very wide variety of audiences including philosophy departments, theology schools, high schools and various physics groups. This talk included some other examples of how fluids ``naturally'' developed complex motions.  An important part of the argument is a demonstration that the laws of fluid motion are not mysterious, but follow directly from a molecular picture of a fluid.  

I gave this talk as a professor of physics and also in my role as president of the American Physical Society.  The point of the talk was not only that fluid flow is an interesting subject.  As a scholar I felt I had a duty to participate in the public discussion on the issues of the day.
The basic  idea is that fluid-motion, while it can be vastly complex,  can be understood by direct application the laws of physics.   
\begin{figure}[!]
 \begin{subfigure}[b]{0.3\textwidth}
                 \includegraphics[width=\textwidth]{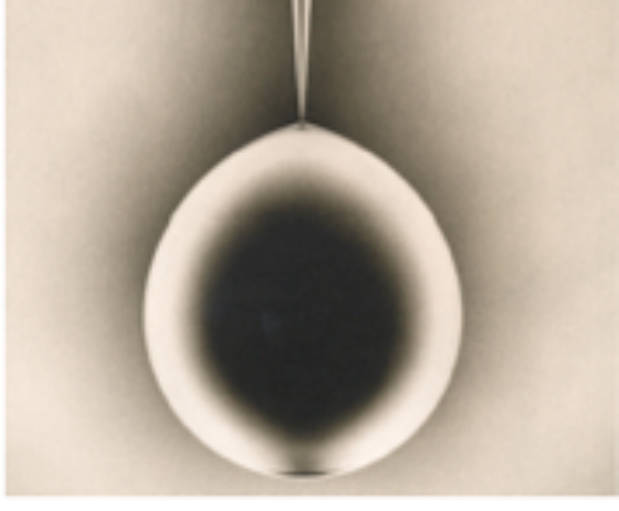}
                 \caption{Drip of Fluid}
                 \label{Drip}
         \end{subfigure}%
%\begin{centering}
%\begin{multicols}{2}
%\beg   in{center}
 \begin{subfigure}[b]{0.6\textwidth}
                 \includegraphics[width=\textwidth]{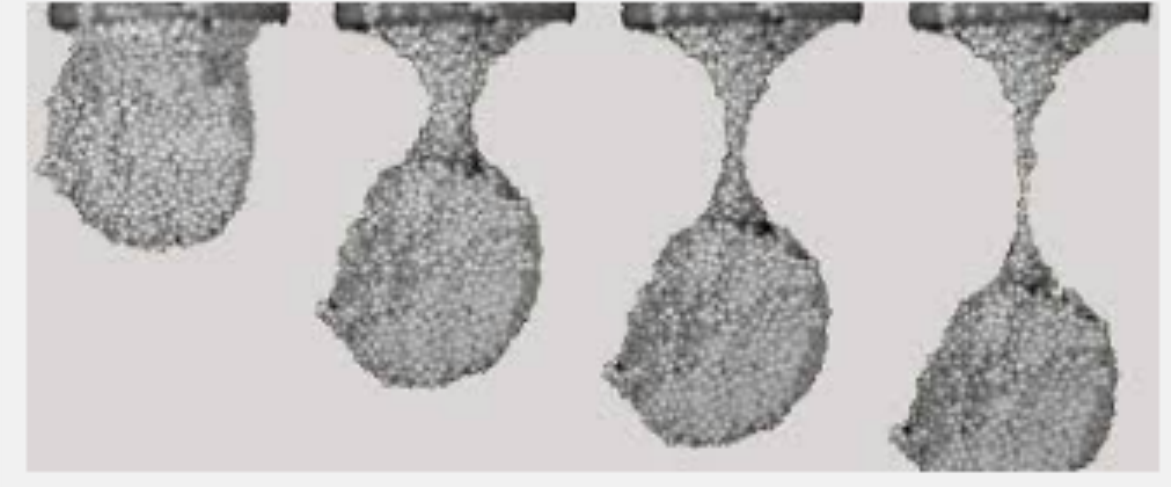}
                 \caption{Fluid containing granular material}
                 \label{GranularDrip}
         \end{subfigure}%

\caption{ Drips.  The first panel, due to Itai Cohen and  Nagel, shows a thin neck just about to break.  The neck is formed by a drop of fluid coming off of a larger mass that has moved through a pipette. The second panel due to Heinrich Jaeger and coworkers\cite{Jaeger} shows a drop composed of a fluid loaded with a granular material.  It too form a drop and a neck, but the detailed shapes in the two panels are entirely different. These flows and shapes  fall into different universality classes.  }
\la{DripFigures}
\end{figure}
But complexity is a fraught word in US science and culture.   That is because there is a large body of people who refuse to believe that humankind is a result of an evolutionary process, but instead argue that we are a product of ``intelligent design'' by a creator or more likely ``Creator.''   Many of them would like this view to become a regular part of school curricula.  Among the people who have contributed to this line of thought are
Reverend William Paley (1743-1805),
Michael Behe (a chemist),
Bill Dembski (in mathematics and philosophy)\footnote{My former student}.  The best argument for intelligent design comes from the Reverend Paley who argued that if someone found a watch on a beach that person would immediately conclude, in view of the careful design and workmanship, that the watch had been designed for a purpose.  Thus, in this line of argument, living things that are even more intricate and involved than the timepiece were probably also created by an intelligent being to serve some purpose.

In our era, 
Professor Behe wrote that living cells are amazingly complex machines    (true).

One cannot imagine how they could have arisen through any natural process.                (He cannot;  I cannot.)

Behe concluded that the only reasonable thing is to assume they were put in place by an intelligent force.

My reply: 
Fluids are basically very simple, they can be described as equivalent to billiard balls in motion,  occasionally bump into one another.
Nonetheless, if you apply heat, they naturally 
and spontaneously form a convective cell, an object resembling a simple machine.   Thus we can see a natural start toward the possible formation of a complex machine.   Perhaps a cascade of such processes might have produced a living cell. Perhaps.

\subsubsection{Knowledge and ignorance}
We don't have much knowledge about how complexity arises, in the present biological world or for the earliest biological forms. Dembski and Behe do us some good by pointing out that ignorance, but neither they nor the Intelligent Design people generally seem to have much positive to add to the discussion. 

My own view is closer to that of Saint Augustine of Hippo  (354-430) 
\begin{quotation}
The Universe was brought into being in a less than fully formed state, but was gifted with the capacity to transform itself from unformed matter into a truly marvelous array of structure and life forms.
\end{quotation}
translation by Howard J. van Till.

\subsubsection{About school curricular}
So, my complexity talk ends with
\begin{quotation}
In the last 1600 years, we have learned some specifics to help flesh out Augustine's good overview.   We should not let the particular view of the universe provided by Intelligent Design's followers replace science in the school curricula.  Instead, we should notice that  
evolutionary biology, paleontology, and cosmology are not speculations. They are root parts of science and human knowledge. They belong in schools as part of the basics of our curriculum and of our understanding of the material world.
Other subjects, unrooted in experiment or scientific observation, should be avoided in school science classrooms. 
\end{quotation}

\subsection{The American Physical Society}
I was proud to serve as President of the American Physical Society (APS), an organization composed of almost 50,000 physicists from the US and the rest of the world.   However, in the three years I was an officer of that group, it seemed to be burdened by a huge inertia.  I wanted the organization to move in directions different from its past history in which it devoted itself to managing its stellar journals, running huge scientific meetings, and lobbying for more government spending on research.   I argued that, in addition, we should have a major effort aimed at the improving scientific education for all US students.  Only slow progress was made in this direction.  I also felt that the APS should be a voice for physicists worrying about tenure, job conditions, and remuneration.  Instead, the APS seemed to be rooted in its past when it spoke for the employers of physicists.

In the years since my APS presidency, the organization has indeed deepened and broadened its educational effort.   It has spoken out against the assassination of scientists in Iran.    So there has been some movement in the right direction.

\subsection{Different kinds of service}
Another one of my ``service'' roles was to serve two terms as director of the University of Chicago's materials research laboratories.   These NSF funded laboratories have as their goal the fostering of studies of condensed matter by diverse teams of scientists.  I found my role immensely satisfying.   I could help my university move in new scientific directions without getting much involved in bureaucratic nonsense. I could foster work on education and help my colleagues make contact with local schools and science museums, thereby building our educational outreach. 

In the same vein,  for about a decade, I managed a program aimed at bringing graduate students into contact with science museums. During five years, we brought about a dozen students each year, half in the sciences and half in the social sciences,  to classrooms where they learned about science museums and to the museums themselves where they helped design exhibits.  We found scientists and museum professionals to meet with the students and spend one afternoon a week teaching them about museums.  The aims of the program included
\begin{itemize}
\item To teach the students about museums and prepare them for museum outreach and museum careers in their future professional life.
\item To bring additional up-to-date science to the museums.
\item To teach the graduate students scientific communication skills.
\item To make university people and museum people more aware of the possibilities of working together.    
\end{itemize}
The program worked quite well for about five years.    But, after a time,  NSF funding for the program disappeared and support for science museums across the country weakened.     At that point, my attention was turning to the APS and with reluctance I gave up that particular museum activity.  

I have been happy to see continued museum outreach within and around our present materials lab.  I still serve this lab, now as ``director'' of outreach activities.  I see my materials lab colleagues and their work on granular materials projected 25 feet high on a wall in Chicago's Museum of Science and Industry.  By the opposite wall I see exhibits for which the early design came from my museum graduate students.      A few of the grad students from the program have begun museum careers.  Others continue to work on presenting science to the public.

\subsection{Scientific criticism}
From a very early point in my professional career, I have worked on a critical assessment of scientific activity.  Many students of the sciences have taken the point of view that science is uniformly wonderful, that all is progress, and that great scientists are always great. A few others have taken the opposite tack.   I have tried to have a more balanced view.  

Among the most recent work is a set of papers about the history of science.  In addition to some laudatory assessments, I have noted that in his first paper on the relevant Fourier series J. Willard Gibbs\cite{LPKGibbs} failed to notice the phenomenon now called the ``Gibbs overshoot.''   In a more serious vein, an assessment of Landau's and BCS's work on the superfluidity\cite{LPKSlippery} indicates that these scholars' important advances in our understanding of microscopic behavior of superfluids and superconductors was matched by an almost complete lack of understanding of the macroscopic wave functions that drive superfluidity.   

My most serious criticism, however, has been devoted to large-scale computer models.  As early as my Urbana days, I started with an assessment of Jay Forrester's {\em Urban Dynamics}\cite{Forrester} model.  This was a set of difference equations which modeled the economic and social growth of an urban area.  He applied a range of different public policy initiatives within the model, and examined their various consequences, saying some initiatives were better and some were worse.   Finding his conclusions in disagreement with my political prejudices, I formed a group that showed how one could reach the opposite ``conclusions'' by changes in the focus of the model and in evaluation criteria.  

In the same vein I studied the use of very large models to simulate the effects of rather turbulent hydrodynamic behavior, not only in convective cells, but also in studies of sonoluminescence and supernova explosions.  (See Fig. (1).)  The results were delivered in papers\cite{LPKExcellence} and in a frequently delivered seminar on ``Excellence in Scientific Computing.''  These models were only partially calibrated, and only included a portion of the physics of the phenomenon at hand.   I concluded with the statement that partially calibrated models were a form of argumentation, and then
\begin{quotation}
How does the power of argumentation provided by exploratory simulations compare to that of rhetorical or order-of-magnitude discussions? Since the simulations must include everything to make a star go boom, they provide an internal check of consistency and completeness not available through words. On the other hand, some intermediate steps in the argument may have their weaknesses hidden in unexamined computer processes. Words may be better than computer output for showing up weak arguments. Computer arguments often force us to rely upon the integrity and care of the investigators. So computers provide a useful but dangerous tool for the exploration of complex systems.

\end{quotation}

Many people including my students,  my postdocs and myself  have done some very well controlled simulations on smaller computers and thereby exposed the qualitative properties of well-defined systems.   When one can guarantee that the results of a computer simulation accurately reflect what will happen in  a specific sharply defined situation, they can be quite meaningful. If, in addition, they answer a well-posed scientific or technical question they are of some value. Otherwise they provide nothing more than a scenario for some imagined event.  

As I see it, very few of the largest and most expensive simulation models answer well-posed questions. In addition, most of these models give results that  should be taken with a grain of salt. 

\section*{Acknowledements}
I want to thank Betsy Kadanoff, Michal Ditzian, and Jim Langer for helpful advice.  This work was partially supported by the University of Chicago NSF-MRSEC under grant number DMR-0820054.

\end{document}